\documentclass[12pt]{article}
\usepackage{a4}
\usepackage{psfig}
\usepackage{epsfig}
\usepackage{epsf}
\usepackage{rotating}
\usepackage{citesort}
\usepackage{amssymb}
\usepackage{graphicx}
\newcommand{\vecc}[1]{\mbox{\boldmath $#1$}}
\newcommand{\vR}{\vecc{R}}            

\newcommand{\matr}[1]{\mbox{#1}}
\newcommand{\eV}{\matr{eV} }
\newcommand{\MeV}{\matr{MeV} }

\def\CCSNO{\mbox{ CC-SNO} }
\newcommand{\boron}{\mbox{${}^8$B\ }}
\newcommand{\fluxunit}{\mbox{$10^6$ cm$^{-2}$ s$^{-1}$ }}

\def\npb#1#2#3{    { Nucl. Phys. }{\bf B #1} (19#2) #3}
\def\npbps#1#2#3{  { Nucl. Phys. }(Proc. Suppl.){\bf B #1} (19#2) #3}
\def\plb#1#2#3{    { Phys. Lett. }{\bf B #1} (19#2) #3}
\def\prd#1#2#3{    { Phys. Rev. }{\bf D #1} (19#2) #3}

\def\prl#1#2#3{    { Phys. Rev. Lett. }{\bf #1} (19#2) #3}

\begin{document}
\begin{flushright}
{\small
CERN-TH-2001-370\\
IFUM-700/FT\\
FTUAM-01-987
}
\end{flushright}
\begin{center}
{ \Large \bf Global analysis of Solar neutrino oscillation evidence
including SNO and implications for  Borexino}\\[0.2cm]

{\large P.~Aliani$^{a\star}$, V.~Antonelli$^{a\star}$,
M.~Picariello$^{a\star}$, 
E.~Torrente-Lujan$^{abc\star}$
\\[2mm]
$^a$ {\small\sl Dip. di Fisica, Univ. di Milano},
{\small\sl and INFN Sezione di Milano, \\ Via Celoria 16, Milano, Italy}\\
$^b$ {\small\sl Dept. Fisica Teorica C-XI, 
Univ. Autonoma de Madrid, 28049 Madrid, Spain,}\\
$^c$ {\small\sl CERN TH-Division, CH-1202 Geneve}\\
}

\end{center}

\begin{abstract}
An updated analysis of all available neutrino oscillation evidence 
 in Solar experiments including the latest $SNO$ data is presented. 
Predictions for total rates and day-night asymmetry in Borexino are 
 calculated.
Our analysis features the use of exhaustive computation of the
 neutrino oscillation probabilities and of an improved
 statistical $\chi^2$ treatment.

In the framework of two neutrino oscillations we conclude that 
 the best fit to the data is obtained in the LMA region with parameters 
 $(\Delta m^2,\tan^2\theta)
  = (5.4 \times 10^{-5}\ \eV^2, 0.38)$, 
 ($\chi^2_{min}/n=0.81$, $n=38$ degrees of freedom).
Although less favored, solutions in the LOW and VAC 
 regions are still possible with a reasonable statistical significance. 
The best possible solution in the SMA region gets a maximum  statistical 
significance as low as $\sim 3\%$.

We study the implications of these results for
 the prospects of Borexino and the possibility of 
 discriminating between the different solutions.
The expected normalized Borexino signal is $0.63$
 at the best fit LMA solution, where the DN asymmetry is 
 negligible (approximately $10^{-2}$).
In the LOW region the signal is in the range  
 $\sim 0.6-0.7$ at $90\%$ confidence level
 while the asymmetry is $\simeq 1-20\%$.
As a consequence, the 
 combined Borexino measurements of the 
 total event rate with an error below $\pm 5-10\%$ and day-night total 
 rate asymmetry with a precision comparable to the one of
 SuperKamiokande will have a strong chance of selecting  or at least 
 strongly favoring one of the Solar neutrino
 solutions provided by present data. 

\vspace{0.5cm}
{\scriptsize \noindent 
$\star$ email: paul@lcm.mi.infn.it, vito.antonelli@mi.infn.it,
 marco.picariello@mi.infn.it, torrente@cern.ch}
\end{abstract}

\newpage

\section{Introduction}

The Solar neutrino
 problem~\cite{homestake,Fukuda:1999ua,Fukuda:1999rq,sage1999,langacker1,RAMOND2,wilczek1,bilenky98,torrentereview,Aliani:2001ba}
 has been defined, for example, 
as the difference between the neutrino flux measured
 by a variety of Solar neutrino experiments and the predictions of the
 Standard Solar Model (SSM).
The explanation of 
this difference  by the neutrino 
oscillation hypothesis has been boosted 
 by the $SNO$ demonstration 
at the $3\sigma$ level 
of the appearance at the Earth 
 of Sun-directed muon and tau neutrino beams~\cite{fiorentini} and the 
 first determination of the total \boron neutrino flux generated
 by the Sun~\cite{sno2001}. 
The agreement  of this flux with the expectations 
 implies as a by-product the confirmation of the validity of the
 SSM~\cite{turck,bpb2001,bp95}.

At present the $SNO$ experiment measures the \boron Solar neutrinos
 via the reactions~\cite{SNO,Boger:2000bb,Barger:2001pf,Bahcall:2001hv,Bahcall:2001hv}:
1) Charged Current ($CC$): $\nu_e + d\rightarrow 2 p+e^-$,
2) Elastic Scattering ($ES$):
 $\nu_x + e^-\rightarrow \nu_x+e^-$.
The first reaction is sensitive exclusively to electron neutrinos.
The second, the same as the one used in SuperKamiokande ($SK$),
 is instead sensitive, with different efficiencies, to all flavors. 
The first results by $SNO$ on Solar neutrinos 
(1169 neutrino events collected during the 
first 240 days of data, Ref.\cite{sno2001}) 
 confirm previous evidence from $SK$ and other
experiments~\cite{homestake,Fukuda:1999rq,Fukuda:1998fd,Fukuda:1998fd,sage1999}.
The Solar neutrino flux measured via the observed CC reaction rate,
 $\phi_{CC}^{SNO}=1.75\pm 0.07\pm 0.12 \times 10^6$ cm$^{-2}$ $s^{-1}$,
 is lower than the two existing ES reaction rate measurements:
 the more precise at $SK$ 
 $\phi_{ES}^{SK}=2.32\pm 0.03\pm 0.08 \times 10^6$
 cm$^{-2}$ $s^{-1}$ and the one measured by  $SNO$ itself
 $\phi_{ES}^{SNO}=2.39\pm 0.34\pm 0.15 \times 10^6$ cm$^{-2}$ $s^{-1}$.
Comparison of the $\phi_{CC}^{SNO}$ and $\phi_{ES}^{SNO}$ and especially
 $\phi_{CC}^{SNO}$ and $\phi_{ES}^{SK}$ 
 can be considered as the first direct evidence for Solar neutrino 
 oscillations which cause the 
 appearance of sun-directed muon and tau neutrinos at the Earth.
When considering only $SNO$ results,
 it is advantageous  to use the experimental ratio of CC to ES measurements
 because of the cancellation of systematic errors.
The experimental value is 
$$
 R^{\rm exp}=\phi_{CC}^{SNO}/\phi_{ES}^{SNO}=0.748\pm0.13
$$
 which is 2$\sigma$ away from the no-oscillation expectation $R^{\rm exp}=1$.
Recall that the statistical significance of the difference 
 $\phi_{CC}^{SNO}-\phi_{ES}^{SNO}$ is clearly lower
 ($\sim 1.4\sigma $)~\cite{sno2001}.
If we use the $ES$-$SK$ result, the no-oscillation hypothesis is excluded at 
 $\sim 3.3\sigma$~\cite{berezinsky2001,poon,torino}.
Moreover the allowed oscillations to {\em only} sterile neutrinos
 is excluded at the same level ($\sim 3\sigma$).
Although less favored,  the oscillations to both active and sterile neutrinos
 are still allowed~\cite{Barger:2001zs,Fogli:2001vr,Krastev:2001tv,Bahcall:2001zu,Bandyopadhyay:2001fb,Choubey:2001bi,Bandyopadhyay:2001aa}. 

Although of qualitative importance, 
 the $SNO$ results by themselves, due to their large error bars, are still 
 far from constraining the neutrino oscillation parameters 
 in a significant way.
For this purpose we continue to need 
 the data of all other Solar neutrino
 experiments~\cite{homestake,Fukuda:1999ua,sage1999,gno2000,gallex}.
The aim of this work is to present an up-to-date analysis of all 
 available Solar neutrino evidence including the latest 
 \CCSNO data in a two-neutrino framework. 
The predictions  for day and night signal rates in the near 
 future real-time ${}^7$Be sensitive Borexino
 experiment~\cite{Hagner:2001sc,Bilenky:2001tf,Arpesella:2001iz,Meroni:2001zj}
 are also calculated. 
Two important characteristics of this paper are the  use a thorough
 numerical computation of the neutrino oscillation probabilities and the
 use of an improved statistical $\chi^2$ analysis. The latter is 
 explained in detail 
in the next section. In summary 
we have followed the $SK$ standard procedure as close as possible 
allowing for a better
 treatment of correlated uncertainties.

As we show below, the best fit to the full data set 
 including global rates and $SK$ energy spectrum is obtained in 
 the {\em Large Mixing Angle} (LMA) region. 
Although less favored, solutions in the {\em Low mass} (LOW) and
 {\em Vacuum} (VAC) 
 regions are possible with a reasonable statistical 
 significance.
A best fit solution in the {\em Small Mixing Angle} (SMA) region is strongly
 disfavored.

This work is organized as follows.
In Section~{\bf\ref{sec:methods}} we present a summary of the methods
 that are used:
 details of the computation of neutrino oscillation probabilities in the 
 Sun and in the Earth, scattering cross sections and calculation of the 
 signal at the detectors.
We continue with a detailed account of the $\chi^2$ statistical methods.
In Section~{\bf\ref{sec:results}} we present the results of different
 analyses: 
 first considering total event rates only and then including
 the $SK$ energy spectrum information.
In Section~{\bf\ref{sec:borexino}} we present our Borexino predictions 
 for the best fit solutions obtained in the previous sections.
Finally in Section~{\bf\ref{sec:conclusions}} we summarize and draw
 conclusions.

\section{Methods}\label{sec:methods}

\subsection{Neutrino oscillations and expected signals}

In general, our determination of neutrino oscillations in Solar
 and Earth matter and of the expected signal in each experiment  follows
 the standard methods found in the literature~\cite{torrente}. 
Nonetheless, this work differs from previous ones  
 in the treatment of the computation of the neutrino transition 
 probabilities.
We completely solve numerically the neutrino equations of evolution for all 
 the oscillation parameter space.
Although other alternatives, as the use of semi-analytical
 expressions in portions of the $(\Delta m^2,\theta)$ plane,
 could be previously justified in terms of economy of resources,
 at this stage a full simulation is an affordable  and better option.
This is specially true  in the calculation of the smearing
 of the neutrino probability as a function of the neutrino production
 point when introducing non radial propagation. However no suprises are 
 found: our numerical 
results are in overall good agreement with the semi-analytical approach.

The survival probabilities for an electron neutrino, produced in the 
 Sun, to arrive at the Earth  are calculated in three steps.
The propagation from the production point to Sun's surface is computed 
 numerically in all the parameter range 
 using the electron number density $n_e$ given by the 
 BPB2001 model~\cite{bpb2001} averaging over the production point.
The propagation in vacuum from the Sun surface to the Earth is computed 
 analytically. The averaging over the annual  variation of the orbit
 is also exactly performed using simple Bessel functions.
To take the Earth matter effects into account, we adopt a
 spherical model of the Earth  density and chemical composition.
In this model, the Earth is divided in eleven  radial density
 zones~\cite{earthprofile}, in each of which  a polynomial
 interpolation is used to obtain the electron density.
The glueing of the neutrino propagation in the three different 
 regions is performed exactly using an evolution operator
 formalism~\cite{torrente}.
The final survival probabilities are obtained from the corresponding
 (non-pure) density matrices built from the evolution operators in 
 each of these three regions.
The  night quantities are obtained using appropriate weights 
 which depend on the neutrino  impact parameter and the 
 sagitta distance from neutrino trajectory  to the Earth center,
 for each detector's geographical location.

The expected signal in each detector is obtained by convoluting neutrino 
 fluxes, oscillation probabilities, neutrino cross sections and detector 
 energy response functions. We have used 
 neutrino-electron elastic cross sections which include radiative
 corrections~\cite{sirlin1994,Passera:2000ug}.
Neutrino cross sections on deuterium needed for the computation of the 
 \CCSNO measurements are taken from~\cite{nakamura2001}.

Detector effects are summarized by the respective response functions,
 obtained by taking into account both the energy resolution and the detector
 efficiency.
The resolution function for \CCSNO is that given in~\cite{sno2001}.
We obtained the energy resolution function for $SK$ using the data
 presented in~\cite{Nakahata:1999pz,skthesis,sakurai}. 
The effective threshold efficiencies, which take into account the live 
 time for each experimental period, are incorporated into our
 simulation program. They are obtained from~\cite{Fukuda:2001nj}. 

\subsection{The $\chi^2$ calculations: definitions and procedures}
\label{section:chi2}

The statistical significance of the neutrino oscillation hypothesis
 is tested with a standard $\chi^2$ method which we explain in some
 detail here.

In the most simple case, the analysis of global rates presented in 
 Section~{\bf\ref{globalsection}}, the definition of the $\chi^2$
 function is the following:
\begin{eqnarray}
  \chi^2_{\rm glob}&=& ({\vR^{\rm th}-\vR^{\rm exp}})^T 
\left (\sigma^{2}\right )^{-1} ({\vR^{\rm th}-\vR^{\rm exp}})
\label{chi1}
\end{eqnarray}
 where $\sigma^2$ is the full covariance matrix made up of two terms, 
 $\sigma^2=\sigma^2_{\rm unc}+\sigma^2_{\rm cor}$. 

The diagonal matrix $\sigma^2_{\rm unc}$ contains the theoretical,
 statistical and uncorrelated errors while $\sigma^2_{\rm cor}$ contains 
 the correlated systematic uncertainties.
The ${\vR}^{\rm th,exp}$ are vectors containing
 the theoretical and experimental data normalized to the SSM 
 expectations.
The length of these vectors is 3 or 4 depending on whether 
 the \CCSNO experiment is included or not: 
$$
R^{th,exp}_i=S_i^{th,exp}/S_i^{SSM},\quad i=Cl,\ Ga,\ SK,\ (\mbox{\CCSNO})\,.
$$
The index $i$ denotes the different Solar experiments:
 Chlorine ($Cl$), Gallium ($Ga$), SuperKamiokande ($SK$) and
 Charged Current $SNO$ (\CCSNO).

The correlation matrices, whether including the $SNO$ experiment or not,
 have been computed using standard techniques~\cite{lisimatrix,Goswami:2000wb}.

To test a particular oscillation hypothesis $(\Delta m^2,\tan^2\theta)$ 
 against the parameters of the best fit, we perform a minimization
 of the $\chi^2_{\rm glob}$ as a function of the oscillation parameters. 
A point in parameter space $(\Delta m^2,\tan^2\theta)$ 
 is allowed if the globally subtracted $\chi_{\rm glob}^2$ fulfills the condition 
 $\chi_{\rm glob}^2 (\Delta m^2, \theta)-\chi_{\rm min}^2<\chi^2_n(CL)$.
Where $\chi^2_{n=2}(90\%,95\%,...)=4.60,5.99,...$  are the $n=2$
 degrees of freedom quantiles.

A more elaborated statistical analysis is necessary once one introduces
 the $SK$ energy spectrum into the analysis
 (Section~{\bf\ref{spectrumsection}}). 

The procedure assumed by the $SK$ collaboration, in order to present its
 own analysis and obtain the tables of bin correlated errors, 
 is that of using the summation $\chi^2=\sum \chi^2_i$ where each  
 term is of the form~\cite{Fukuda:2001nk,Fukuda:2001nj}
\begin{eqnarray}\label{cor1}
 \chi_i &\sim & \alpha \ f_i(\delta_{\rm cor})\  R_i^{\rm th}-R_i^{\rm exp},
\end{eqnarray}
 and where the expression
\begin{eqnarray}\label{cor2}
 f_i(\delta_{\rm cor})&\sim & \frac{1}{1+\sigma_{i,{\rm cor}}\delta_{\rm cor}}
\end{eqnarray}
 is the response function for the  correlated error in the
 $i^{th}$-energy bin. 
The $\sigma_{i,{\rm cor}}$ are bin-correlated uncertainties and $\alpha$  
 is an overall flux normalization factor.
The correlation parameter $\delta_{\rm cor}$ is arbitrary
 and determined in the minimization 
 process together with the rest of oscillation parameters. 

SuperKamiokande calculation of  tables for correlated errors 
 (as Table III in Ref.~\cite{Fukuda:2001nj} and previous publications)
 and the  $\chi^2$ prescription given respectively by
 Eqs.~(\ref{cor1}) and~(\ref{cor2}) are closely related.
In this work we follow this definition of the $\chi^2$, however 
 for clarity we write the expressions in a notation
 which renders the comparison with other published works simple. 
It can be easily shown that the expression to be presented below
 leads to the $SK$ expressions~\footnote{We have not included a term 
 $\log\det \sigma^2(\delta_{\rm cor})\sim \log (1+\delta_{\rm cor})$
 whose effect is negligible}.

Therefore, for  the analysis of the full set of data including $SK$ energy
 spectrum, we consider  a $\chi^2$ function which is  sum of two quantities
 $\chi^2=\chi^2_{\rm glob}+\chi^2_{\rm spec}$.
The first one is given by Eq.~(\ref{chi1}) considering only 
 $Cl$, $Ga$ and \CCSNO total rates, while for the 
 second term  we adopt the definition 
\begin{eqnarray}\label{chi2}
  \chi^2_{\rm spec}&=&\sum_{d,n} ({\alpha \vR^{\rm th}-\vR^{\rm exp}})^t 
\left (\sigma^{2}_{\rm unc}+\delta_{\rm cor} \sigma^{2}_{\rm cor}\right )^{-1}
 ({\alpha \vR^{\rm th}-\vR^{\rm exp}})
 +\chi_\alpha^2+\chi^2_\delta\, .
\end{eqnarray}
A subindex $i=d,n$ corresponding to separated day and night 
 quantities is understood for any $\vecc{R}$ vector and 
 $\sigma$ matrix.
We have introduced the flux normalization factor $\alpha$ and the correlation 
 parameter $\delta_{\rm cor}$.
The complete variance matrix is not a constant quantity.
It is obtained by combining the statistical variances with systematic
 uncertainties and dependent on this correlation parameter.

The corresponding $\chi^2$ terms in Eq.~(\ref{chi2}) are 
$$
\chi^2_\alpha =
 \frac{\left(\alpha - \alpha^{\rm th}\right)^2}{\sigma_\alpha^2},\quad 
\chi^2_\delta=
 \frac{\left(\delta_{\rm cor} - \delta_{\rm cor}^{\rm th}\right)^2}
        {\sigma_\delta^2}\,.
$$
The central value $\alpha^{\rm th}=1$ corresponds to the 
 BPB~2001 \boron flux.
The uncertainty on $\alpha$ comes from the ${}^{+19}_{-14}\%$ 
 uncertainty in the \boron flux and we make an average of the 
 two-side errors taking
 $\alpha^{\rm th}\pm \sigma_\alpha=1.00\pm 0.17$~\cite{bpb2001}.
The correlation parameter is assumed to be constrained 
to vary in a gaussian way with an error corresponding to
 $\sigma_\delta$ and we take $\delta_{\rm cor}^{\rm th}=1$.

In this work we consider two cases. In the first case the flux
 normalization is taken to be free (we allow for a very large
 $\sigma_\alpha\to \infty$ in the $\chi_\alpha^2$ term).
As a second possibility we consider that both quantities $\alpha$ and
 $\delta_{\rm cor}$ are constrained. 
Note that the unwanted consequence of eliminating the last term
 $\chi^2_\delta$ would be to obtain running-away solutions for
 the optimal value of this parameter.

The $\chi^2$ summation now contains 41 bins in total: 3 from
 the global rates (all the experiments except $SK$)  
 and 2$\times$19 bins for the $SK$ day and night spectrums. 
The full correlation matrix is defined by blocks. 
The  3$\times$3 block corresponding to the global rates is defined
 as above.
For each day and night spectrum the corresponding 19$\times$19 block
 correlation matrices are conservatively constructed assuming full
 correlation among energy
 bins~\footnote{The introduction of the parameter $\delta_{\rm cor}$ is
 equivalent to the relaxation of this condition during minimization.}.
The components of the variance matrix are 
\begin{eqnarray}
 \left (\sigma_{\rm cor}^2 \right )_{ij}&=& 
 \sigma_{i,\rm exp}\sigma_{j,\rm exp}+\sigma_{i,\rm cal}\sigma_{j,\rm cal},
\nonumber \\
\left (\sigma_{\rm unc}^2 \right)_{ii}&=& \sigma_{i,\rm stat}^2
 + \sigma_{i,\rm unc}^2, 
\nonumber
\end{eqnarray}
 where $\sigma_{\rm stat}$ are the statistic errors and the quantities 
 $\sigma_{\rm exp},\sigma_{\rm cal},\sigma_{\rm unc}$ are  respectively 
 the bin-correlated experimental, the spectrum calculation uncertainties 
 and the bin-uncorrelated ones.
 
Notice that in the above sum we do not simultaneously include 
 the global ratio for $SK$ and the partial spectrum bins. 
We ignore any correlation between the spectrum information and
 the global rates of all experiments except $SK$ itself.
Finally, a remark is in order.
In this case the defined covariance matrix is dependent on one of the
 fitting parameters.
One does not, however have to add any correction
 (i.e. $\log \det \sigma$ terms) to the $\Delta \chi^2$ expression 
 as long as we sit at $\delta_{\rm cor}^{\rm min}$. 
To test a particular oscillation hypothesis and
 obtain allowed regions in parameter space we perform a 
 minimization of the four dimensional function
 $\chi^2(\Delta m^2,\tan^2\theta,\alpha,\delta_{\rm cor})$. 
The minimization of the expression with respect to $\alpha$ and $\delta_{\rm cor}$ 
 is done analytically.
The subsequent minimization in the $\left (\Delta m^2, \tan^2\theta\right )$
 plane is numerical.
For $\delta_{\rm cor}=\delta_{\rm cor}^{\rm min},\alpha=\alpha_{\rm min}$, 
 a given point in the oscillation parameter space is allowed if 
 the globally subtracted quantity fulfills the condition 
 $\Delta \chi^2=\chi^2 (\Delta m^2, \theta)-\chi_{\rm min}^2<\chi^2_n(CL)$.
Where $\chi^2_{n=4}(90\%,95\%,...)=7.78,9.4,...$ are the quantiles for
 four degrees of freedom.

\section{Results}\label{sec:results}

We have used data on the total event rates measured at chlorine Homestake 
 experiment, at the gallium experiments $SAGE$~\cite{sage,sage1999},
 $GNO$~\cite{gno2000} and $GALLEX$~\cite{gallex} and at the 
 water and heavy-water $SK$ (live time 1258 days)~\cite{Fukuda:2001nk}
 and $SNO$ ($ES$,$CC$ 240 days~\cite{sno2001}) experiments
 (see Table~(\ref{tableratios}) for an explicit list of results and 
 references).
For the purposes of this work it is enough to summarize all the 
 gallium experiments in one single quantity by taking the weighted
 average of their rates.

In addition, for $SK$ we use the available information for the day
 and night energy spectrums.
For each of the day and night cases, this spectrum information
 contains 18 total recoil-electron energy bins of width 0.5 \MeV
 in the range 5 to 14 \MeV and an additional
 bin spanning the remaining 14-20 \MeV range.
The data and errors for individual energy 
 bins for $SK$ spectrum has been obtained from Ref.~\cite{Fukuda:2001nj}. 
The information from other $SK$ results as the global 
 day night asymmetry is  already contained to a large 
 extent in the previous quantities and 
 does not change the results to be presented on
 continuation~\cite{Fukuda:2001nk,Fukuda:1999rq}.

\subsection{Global rate analysis}
\label{globalsection}

The analysis of the global rates of the four experiments $SK$, Chlorine, 
 Gallium and \CCSNO is the simplest possibility but nonetheless it  reveals 
 important trends of the solutions to the Solar neutrino problem.
It illustrates how the \CCSNO data impose additional constraints on
 the parameter space and in particular how the SMA solution loses its
 statistical significance in favor of the LMA and LOW regions,
 becoming allowed only at marginal $\approx 3\sigma$ confidence levels.

We present results both with and without the inclusion of the \CCSNO
 global rate.
In the analysis of the global rates of the $SK$, $Cl$, $Ga$ (\CCSNO)
 experiments one has two free parameters $\Delta m^2$ and $\tan^2 \theta$
 and 3 (4) experimental quantities, therefore the effective number of
 degrees of freedom (d.o.f.) is 1 (2).

On the left-hand-side of Tables~(\ref{table1a}-\ref{table1b})
 we present the best fit parameters or local minima 
 obtained from the minimization of the $\chi^2$ function given in 
 Eq.~(\ref{chi1}).
Also shown are the values of $\chi^2_{\rm min}$ per degree of freedom
 ($\chi^2/n$) and the goodness of fit (g.o.f.) or significance level of each 
 point (definition of SL as in Ref.~\cite{pdg}).
On the right-hand-side  of the same tables we show the deviations
 (minimization residuals) from the expected values for all the total
 event rates. 
We also include, for further reference, the deviations corresponding to
 the $SK$ day and night energy spectrums and global 
 day-night asymmetry, although these quantities are not incorporated in the 
 $\chi^2$ minimization.  
We first consider the case in which the \CCSNO global rate is ignored. 
The absolute minimum is located at the SMA region 
 $(\Delta m^2,\tan^2\theta)^{SMA}
 = (7.8 \times 10^{-6}\ \eV^2, 1.10 \times 10^{-3})$,
 with 
 $\chi^2_{\rm min}/n=1.06$ 
 (d.o.f. $n=3-2$), the significance level for this point g.o.f. is $29\%$.
Closely we find the twin 
minima situated in the VAC region with parameters
 $(\Delta m^2,\tan^2\theta)^{VAC}
  = (8.4  \times 10^{-11}\ \eV^2, 3.37 (0.27))$, 
 which receive a significance level only 
slightly smaller  than the SMA mininimum.
The next local minima are situated in the LMA region
 $(\Delta m^2,\tan^2\theta)^{LMA}=(1.23 \times 10^{-5}\ \eV^2, 0.34)$, 
 with  
 $\chi^2_{\rm min}/n=1.89$
 which corresponds to a still reasonable good fit g.o.f. is $17\%$
 and in the LOW region 
 $(\Delta m^2,\tan^2\theta)^{LMA}
  = (1.11 \times 10^{-7}\ \eV^2, 0.75)$, 
 with  
 $\chi^2_{\rm min}/n=3.62$.
For comparison, much-less significant results are obtained when 
sterile neutrino oscillations are considered.

When one introduces the \CCSNO total event rate in the analysis the 
 relative order of the local minima changes.
The absolute minimum is now located at the LMA region:
 $(\Delta m^2,\tan^2\theta)^{LMA}
  = (3.12 \times 10^{-5}\ \eV^2, 0.40)$. 
The significance level of the absolute minimum is clearly  worse
 than in the previous case $\chi^2_{min}/n=1.61$ (d.o.f. n=4-2) which
 corresponds to g.o.f. $=21\%$.

The significance level of the SMA region which follows 
is  no better 
than $\chi^2_{\rm min}/n=2.40$, g.o.f. is $8.8\%$, value which is obtained at 
 $(\Delta m^2,\tan^2\theta)^{SMA}
 = (7.3 \times 10^{-6}\ \eV^2, 1.40 \times 10^{-3})$. 
After that, we have the pair of minima at the VAC region 
 $(\Delta m^2,\tan^2\theta)^{VAC}
 = (9.7\times 10^{-11}\ \eV^2, 4.00 (0.26))$, 
 with
 $\chi^2_{\rm min}/n=3.2-3.1$.
Finally, we find in the table 
a  minimum in the LOW closely situated to the one of the 
previous case.

\begin{figure}[p]
\centering
\begin{tabular}{ll}
\psfig{file=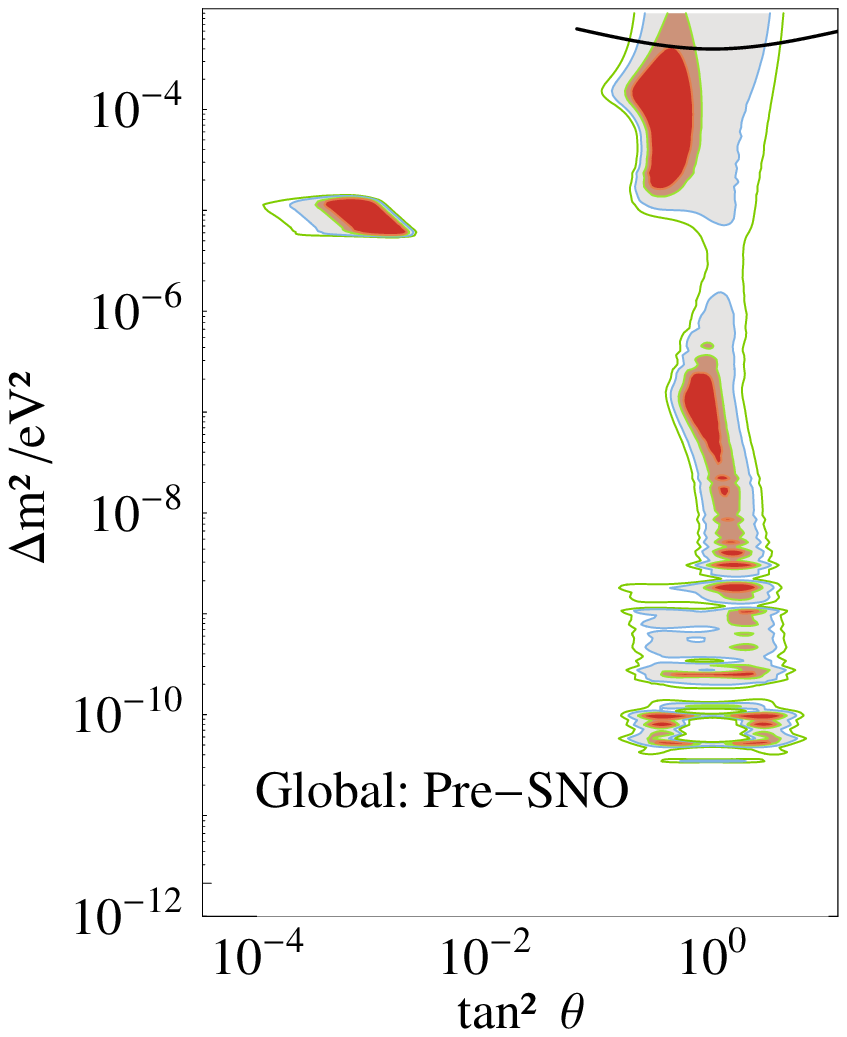,width=8cm}& \hspace{-1.1cm}
\psfig{file=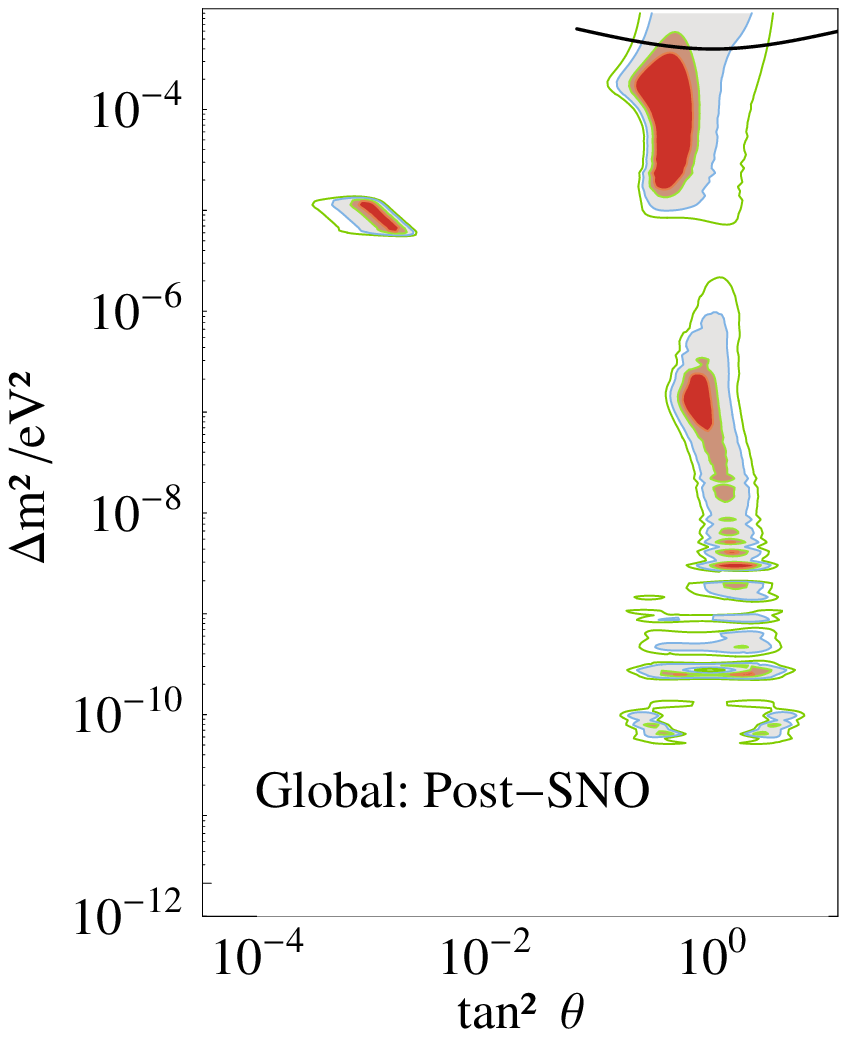,width=8cm} 
\end{tabular}
\caption{\small
Best fit solutions for the global rate analysis.
Global rates from $SK$, $Cl$ and $Ga$ experiments are included.
The right figure contains also the \CCSNO one.
The black dots are the best fit points
 (the absolute minimum is located at the SMA (left) or LMA (right) regions).
The colored areas are the allowed regions at
 90, 95, 99 and 99.7\% CL relative to the absolute minimum.
The region above the solid line is excluded by $CHOOZ$ results at 99\% CL 
\protect\cite{chooznew}.
}
\label{f1a}
\end{figure}

In Figs.~(\ref{f1a}) we present graphically our results respectively
 before and after  including the global \CCSNO rate.
In the plots one can see the regions which are allowed at 
 90, 95, 99 and $99.743\%$ confidence levels.
The main difference between the plots is an overall reduction of
 the extent of the allowed area at any CL.
The large contiguous areas at 3 and $4\sigma$ which are present 
 along $\tan^2\theta\sim 1$ in the first case are converted into 
 well separated patches after the inclusion of $SNO$ results.
The most serious consequence is however the considerable reduction on the
 significance of the SMA region.
From the analysis of the four global rates a good part of this region becomes 
acceptable only at some marginal level.
In these and next plots, the region above the line at high $\Delta m^2$ is 
 excluded at $99\%$ CL from the negative results of $CHOOZ$ and Palo
 Verde~\cite{chooznew}.
Note that regions for very large $\Delta m^2\sim 10^{-3}$ \eV$^2$ 
 are excluded at the $2-3 \, \sigma$ level without need of the $CHOOZ$ result once we included the $SNO$ data.

The reduction of the SMA region can be explained from the
 tables of deviation (right part in Tables~(\ref{table1a}-\ref{table1b})).
Before the introduction of the \CCSNO global rate in the computation and 
 $\chi^2$ minimization  the residuals for the \CCSNO data corresponding to the
 best fit are very high (Table~(\ref{table1a})): 
  $2.5\sigma$ for the SMA and also 
 for the LOW solution.
The \CCSNO residual values for remaining regions are however smaller.  
When the \CCSNO data is incorporated into the fit the SMA and LOW regions 
 become less favored.
The effect is much more important for the relatively small SMA region.
For the LOW region the initial area is large and nearby minima can be found
 to adjust for the \CCSNO data and still provide reasonable quality fits.


\subsection{ Day-night $SK$ spectra and global rates}
\label{spectrumsection}

In this section we present the results obtained when including the 
 $SK$ day and night energy spectrum rates in addition to the total
 event rates of $Cl$, $Ga$ experiments and \CCSNO.

The definition of the $\chi^2$ functions and minimization 
 procedures used in the statistical analysis are explained in detail 
 in Section~{\bf\ref{section:chi2}}. 
 The number of experimental data inputs is  $2\times 19+3=41$.
 One now has four free parameters, the oscillation 
 parameters $\Delta m^2$ and $\tan^2\theta$, the \boron neutrino 
 flux normalization factor $\alpha$ and the correlation parameter
 $\delta_{\rm cor}$.
Two cases will be studied.
In case A the flux normalization $\alpha$ is considered a free parameter, 
 the number of effective d.o.f. is then $41-4=37$.
In case B, the parameter $\alpha$ is constrained to vary around the BPB2001
 central value with a  standard deviation $\sigma$ given by SSM. We now have 
 $38$ d.o.f.

In Tables~(\ref{table2a}-\ref{table2b}) we present the best fit
 parameters or local minima obtained from the minimization of the
 $\chi^2$ function given in Eq.~(\ref{chi2}).
As in the previous section, the values of $\chi^2_{\rm min}$ per degree
 of freedom ($\chi^2/n$) and the goodness of fit (g.o.f.) or
 significance level of all minima are shown.
In the right part of the tables, the deviations 
 from the expected values or minimization residuals for a number of
 experimental quantities are listed.

\begin{figure}[p]
\centering
\begin{tabular}{rl}
\psfig{file=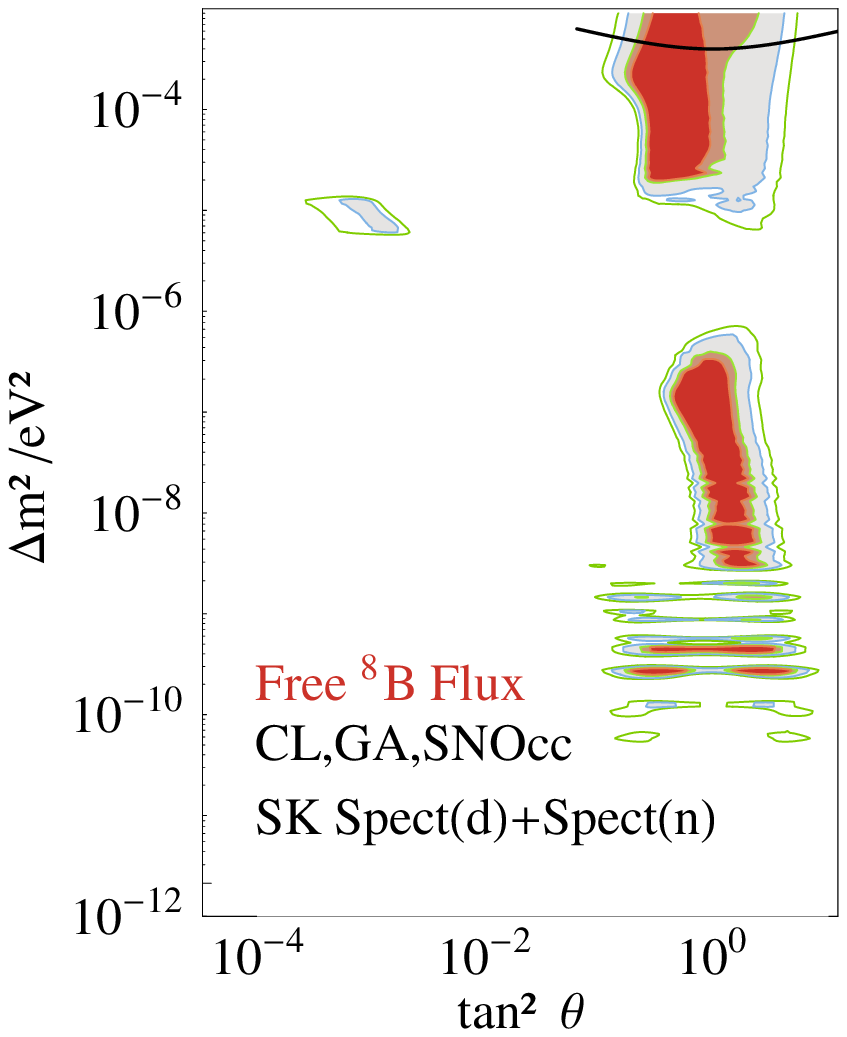,width=8cm}& \hspace{-1.1cm} 
\psfig{file=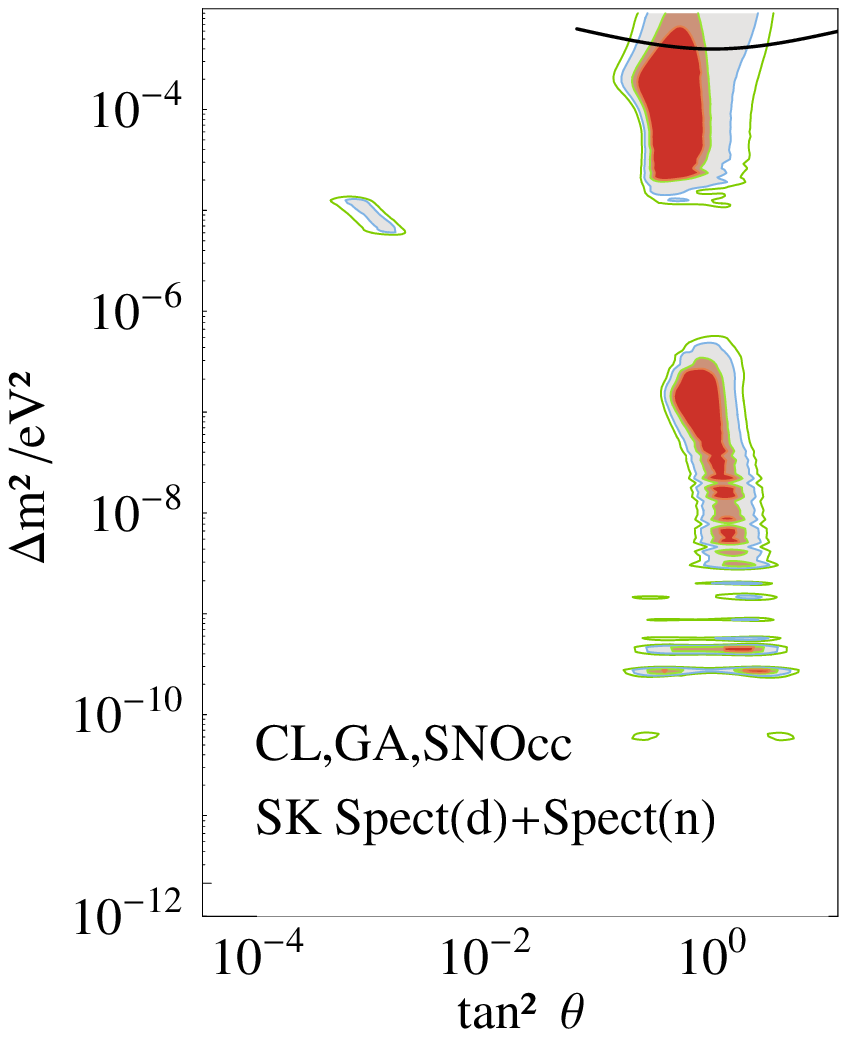,width=8cm} 
\end{tabular}
\caption{\small
Global solutions for the analysis of $Cl$, $Ga$ and \CCSNO experiments
 plus the day-night $SK$ spectrum rates.
The black dots correspond to best fit points.
The colored areas are the allowed regions at 
90, 95, 99 and 99.7\% CL relative to the absolute minimum.
Left (case A of the text):
 the minimization with respect $\delta_{cor}$ is performed as before.
 The absolute normalization factor $\alpha$ is allowed to vary freely.
Right (case B of the Text): 
 the absolute normalization factor $\alpha$ and correlation 
 parameter $\delta_{\rm cor}$ are varied subject to constraints.
}
\label{f2}
\end{figure}

The results for the free flux minimization or case A are presented first.
As a summary of  numerical results, which appear in detail in 
 Table~(\ref{table2a}) and Fig.~(\ref{f2}, left), 
 the position of the absolute minimum is located at the LMA region 
 $(\Delta m^2,\tan^2\theta)^{LMA}=$ 
$ (5.4 \times 10^{-5}\ \eV^2, 0.36)$, 
 with  
 $\chi^2_{\rm min}/n=0.82$ (d.o.f. n=37).
The significance level for this point is noticeably larger 
(g.o.f. $=82\%$) than the  levels obtained in global
 rates only analysis.
This is only in part an artifact of the statistical machinery.
It is a satisfactory result because it basically reflects the internal
 consistency of the data: although the number of degrees of freedom have 
 considerably increased, the $\chi^2$ per d.o.f is still much the same.   
The minimization value of the flux normalization  is $\alpha=1.02$ while 
 the correlation parameter is $\delta_{\rm  cor}=1.43$.
The next local minima are situated at the following positions: the 
 LOW region
 $(\Delta m^2,\tan^2\theta)^{LOW}
  = (1.2 \times 10^{-7}\ \eV^2, 0.75)$, 
 with
 $\chi^2_{min}/n=0.85$ 
 and the two solutions at the VAC region 
 $(\Delta m^2,\tan^2\theta)^{VAC}
  = (2.7 \times 10^{-10}\ \eV^2, 2.70 (0.32))$, 
 with  
 $\chi^2_{\rm min}/n=0.85$.
In this last region the best-fit flux normalization
 is down to $\alpha=0.56$ while $\delta_{\rm cor}=1.65$.
In case B, where the flux normalization is constrained to its SSM value, 
 the results are very similar and are shown in Table~(\ref{table2b})
 and Fig.~(\ref{f2}, right).
The position of the first minimum remains practically unchanged,
and is located at the LMA region 
 $(\Delta m^2,\tan^2\theta)^{LMA}
 = (5.4 \times 10^{-5}\ \eV^2, 0.38)$, 
 with
 $\chi^2_{\rm min}/n=0.82$ 
 (d.o.f. n=38) and g.o.f.$=82\%$.
The next local minima are: the one situated in
 the LOW region
 $(\Delta m^2,\tan^2\theta)^{LOW}
  = (7.5 \times 10^{-8}\ \eV^2, 0.84)$, 
 with
 $\chi^2_{\rm min}/n=0.86$,
 and the twin solutions at the VAC region 
 $(\Delta m^2,\tan^2\theta)^{VAC}
 = (8.9 \times 10^{-10}\ \eV^2, 1.73(0.48))$, 
 with  
 $\chi^2_{\rm min}/n=0.9$.
Finally,
in both cases A and B we have a local minimum situated at the SMA region.
These minima are situated at  $\Delta m^2\simeq 7.3 \times 10^{-6}$ and
 $\tan^2\theta\simeq 1.3 \times 10^{-3}$ with a considerable poorer 
statistical significance.

\begin{figure}[p]
\centering
\begin{tabular}{ll}
\psfig{file=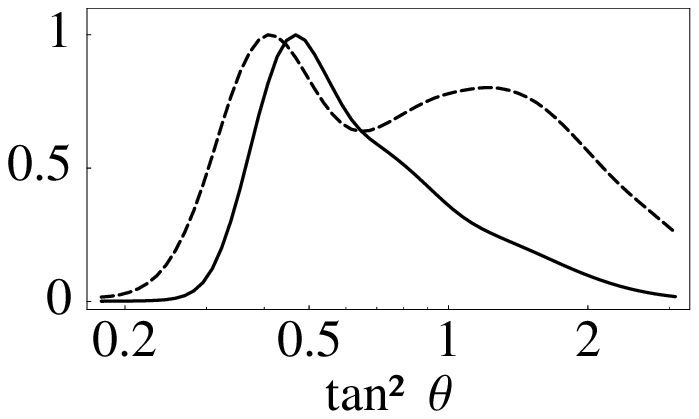,width=8cm,height=6cm} &\hspace{-1.5cm}
\psfig{file=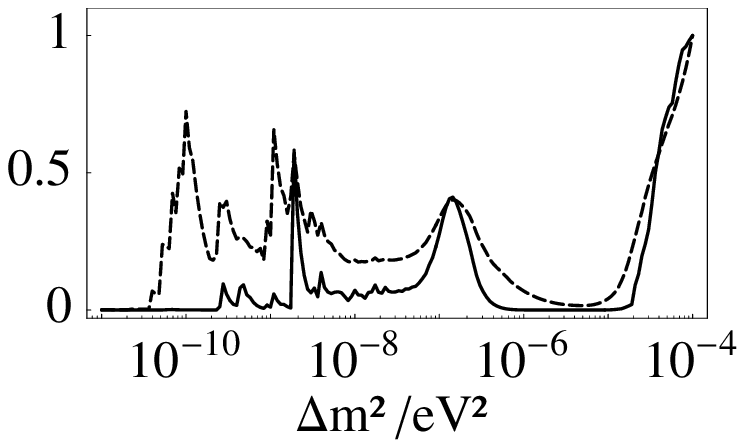,width=8cm,height=6cm} 
\end{tabular}
\caption{\small
Marginalized likelihood distributions for each of the
 oscillation parameters $\Delta m^2$ (right), $\tan^2 \theta$ (left)
 appearing in the $\chi^2$ fit.
The curves are in arbitrary units with normalization to the 
 maximum height.
The continuous lines shows the distributions corresponding to 
 Fig.~(\protect\ref{f2}, right). Dashed lines correspond to the case 
 represented in Fig.~(\protect\ref{f1a}, left). 
}
\label{f6}
\end{figure}

From the Tables~(\ref{table1a}-\ref{table1b}-\ref{table2a}-\ref{table2b}) 
 and also from $\chi^2$ landscape scans it can be 
  seen that there are some non-negligible regions in parameter space,
 not only the best fit points.
We are therefore justified in converting $\chi^2$ into likelihood
 using the expression ${\cal L}=e^{-\chi^2/2}$,
 and proceeding to study the marginalized parameter constraints.
This normalized marginal likelihood is plotted
 in Figs.~(\ref{f6}) for
 each of the oscillation parameters $\Delta m^2$ and $\tan^2\theta$.
We present  results corresponding to two cases: the global-rate analysis
 excluding \CCSNO and the full analysis including $SK$ energy spectrum. 
For $\tan^2\theta$ we observe that the likelihood function is concentrated in a 
 region $0.2<\tan^2\theta<1$ with a clear maximum at $\tan^2\theta\sim 0.5$ in 
 sharp coincidence with previous results.
The situation for $\Delta m^2$ is less obvious although the region 
 at large mass differences is clearly favored.
The half width of these curves can be used as another estimate of 
 the minimum error which can be assigned to the parameters at the
 present and near future experimental situation.

\section{Borexino implications}\label{sec:borexino}

Borexino is a real-time detector for low energy
 ($<1$ \MeV) spectroscopy.
The experiment's goal is the direct measurement of the 
 ${}^7$Be Solar neutrino flux of all flavors via neutrino-electron
 scattering in an ultra-pure scintillation
 liquid~\cite{Hagner:2001sc,Arpesella:2001iz,Meroni:2001zj,Bilenky:2001tf}. 

As before, in order to  estimate the Borexino signal one convolutes the 
 Sun-neutrino flux with a detector response function which includes
 the elastic scattering cross sections 
 $\nu_{e,\mu} e^- \rightarrow \nu_{e,\mu} e^-$,
 energy resolution and energy-dependent efficiency effects. 
We assume a Gaussian resolution function with an energy dependent 
 width $\sigma(E_{exp})=0.048 \, \sqrt{E_{exp}}+ 0.01$ \MeV.
At this stage we suppose unit efficiency over the nominal window of the 
 experiment $0.3< E<0.8$ \MeV and zero otherwise.
The absolute ${}^7$Be flux is taken from BPB2001~\cite{bpb2001} as
 for the other experiments. 
After one year of data taking, Borexino~\cite{private:bor}
 expects roughly \mbox{$15-20\times 10^3$} Solar neutrino events with 
 a nearly negligible statistical error $\sigma_{stat}\sim 0.8 \%$ whereas 
the error in discriminating signal from background
 is much larger $\sigma\sim 8\%$ for the same period.

\begin{figure}[p]
\centering
\psfig{file=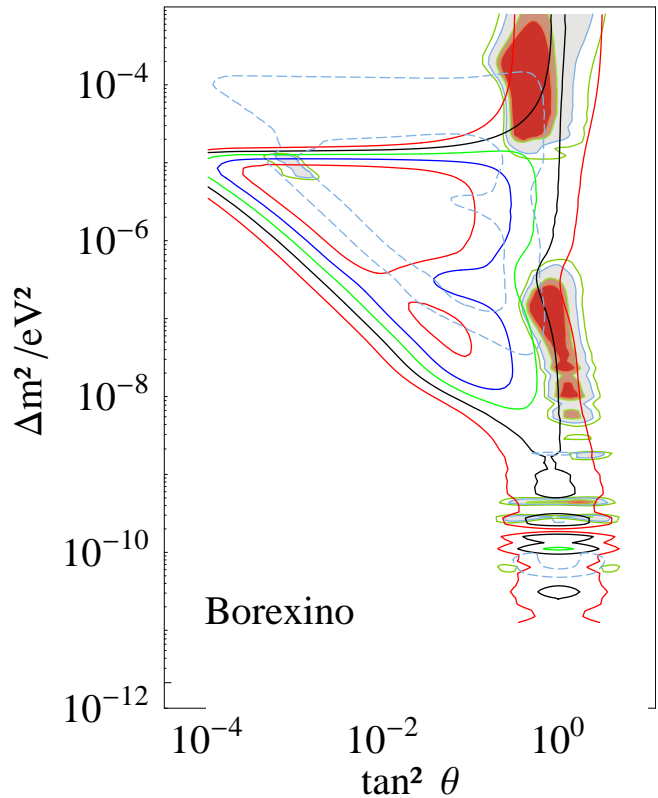,width=10cm}
\caption{\small 
The signal at the Borexino experiment as a function of the 
 oscillation parameters.
The signal is normalized to the no-oscillation case.
Contours are drawn at $S/S_{0}=0.5, 0.6, 0.7$ with full lines
 (respective from inside to outside).
Superimposed: the allowed regions from the global rate analysis
 including \CCSNO (see Fig.~(\protect\ref{f2}), right)
 and the $1\sigma$ allowed regions (here enclosed by the dashed lines)
 from the ${}^8$Be sensitive $Cl$ experiment alone.}
\label{f7}
\end{figure}

In Table~(\ref{table7}) and Fig.~(\ref{f7})
 we present the expectations for the normalized  
 day-night signal $S^{day-night}$ at the Borexino experiment for all of the 
 local minima solutions found in the previous section 
 (SK spectrum plus global rates with constrained flux): 
 $S^{Bor}=S^{D-N}(\Delta m^2,\theta)/S_{0}$ where 
 $S_{0}$ is the expected signal in absence of oscillations 
 and day-night signals are 
 averaged.
We also present the expected day-night asymmetry $A^{DN}\equiv 2 (D-N)/(D+N)$.
As we will see below, the asymmetry on the day-night event rates is a
 valuable tool for distinguishing among the different oscillation solutions. 
In Fig.~(\ref{f7}) we present the expectations 
 for the Borexino experiment as a function of 
 the two-dimensional oscillation parameters.
The expected signal varies mildly in ample regions of the
 parameter space.
We observe that at the best fit solution, situated at the LMA region,
 the  expected Borexino normalized signal is $S^{Bor}=0.63$.
In the whole LMA region (99.7\% CL) the signal varies between 0.5 and 0.7. 
If we restrict ourselves to the 90\% CL allowed region around the absolute 
 minimum the signal is always within the range $\sim 0.6-0.7$.
The asymmetry at the absolute minimum is negligible
 $A^{DN}<10^{-5}$, while in the LMA region which surrounds it,
 it can reach the $\sim 1\%$ level.
A similar behavior appears in all the rest of the parameter 
 space except in the LOW region.

In the LOW region the signal varies between $\sim 0.6-0.7$ (90\% CL). The 
 variation increases to cover the range $\sim 0.5-0.8$ at $99.7$ CL.
Here the day-night asymmetry is expected to be at its maximum.
From the table one observes values as high as $A^{DN}\simeq 30\%$.

\begin{figure}[p]
\centering
\psfig{file=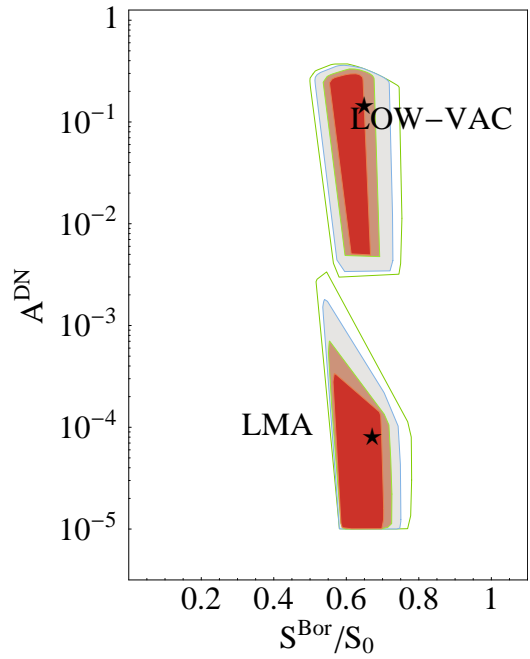,width=8cm}
\caption{\small 
Day night asymmetry $\mid A^{DN}\mid $ versus normalized signal at Borexino corresponding
to the 90, 95, 99, 99.7 \% CL allowed regions from the full analysis including 
global rates and SK spectra.
The upper (lower) zones correspond to the LOW and VAC (LMA) regions in 
Fig.~\protect(\ref{f2}, left). The stars are the expectations 
of the best fit solutions in each of the regions.
}
\label{f8}
\end{figure}

In the VAC region the signal varies between $\sim 0.7-0.8$.
The variation of the expected signal in the SMA region is much more
 abrupt.
This prevents us from giving accurate predictions for this case.
The signal passes
 from a value $S^{Bor}\sim 0.2$ to $S^{Bor}\sim 0.8$ in a very narrow region.
As it is shown in Fig.~(\ref{f8}),
 the combined Borexino measurements of the 
 total event rate with an error below $\pm 7-10\%$ and day-night total 
 rate asymmetry with a precision comparable to that of $SK$ will 
 allows us to distinguish or at least to strongly favor the Solar neutrino
 solutions provided by present data.

\section{Conclusions}\label{sec:conclusions}

We have analyzed experimental evidence from all the 
 Solar neutrino data available at this moment including the latest charged 
 current $SNO$ results. 
For the best  solutions which fit  the present data, 
 we have obtained and analyzed the expectations in Borexino.
We considered different combinations of non-redundant  data to perform 
 two kinds of analyses: 
 in the first one we only included total event rates for all the experiments, 
 while in the second we included global rates 
 for $Homestake$, $GNO$-$GALLEX$-$SAGE$ and  \CCSNO plus the 
 day and night energy spectrum rates provided by $SK$.
In this last case we studied the possibility of free or
 constrained variations of the Boron Solar neutrino flux.

In the simplest case involving global rates only we have observed how the 
 inclusion of the \CCSNO data causes the strong decrease of the 
 statistical significance of the SMA solution.

In the most comprehensive case, global rates plus spectrum,
 the best fit was obtained in the LMA region with parameters 
 $(\Delta m^2,\tan^2\theta)^{LMA}
  = (5.4 \times 10^{-5}\ eV^2, 0.38)$, 
 ($\chi^2_{\rm min}/n=0.82$, d.o.f. n=38).
Solutions in the LOW and VAC regions are  still possible although
 much less favored. 
The best possible solution in the SMA region gets a low statistical
 significance.

We have analyzed in detail the expectations on the future experiment Borexino for all 
the favored  neutrino oscillation solutions.
The expected Borexino normalized signal is $S^{Bor}=0.63$ 
 at the best-fit LMA solution while the day-night asymmetry is 
 negligible $A^{DN}<10^{-5}$.
In the VAC region the signal is slightly higher 
 $\sim 0.7-0.8$ and the asymmetry is still practically negligible.
In the whole LMA 
 region (99.7\% CL) the signal varies between 0.5 and 0.7. 
In the LOW region the signal is in the range  $\sim 0.6-0.7$ at 
 90\% CL while the asymmetry $A^{DN}\simeq 1-30\%$.
We conclude that in the near future, after 2-3 years of data taking, 
 the  combined Borexino measurements of the 
 total event rate with an error below $\pm 5-10\%$ and day-night total 
 rate asymmetry with a precision comparable to that of $SK$ will 
 allows us to distinguish or at least to strongly favor the Solar neutrino
 solutions provided by present data.
Additional information and precision measurement should be obtained also 
from Kamland and from future solar neutrino experiments~\cite{strumia}
\subsection*{Acknowledgments}
It is a pleasure to thank R. Ferrari for many enlightening discussions and 
 for his encouraging support without which this work would have not been 
 possible.
We thank all the Borexino group of Milano University and especially
 M. Giammarchi, E. Meroni and B. Caccianiga for providing us
 essential information about the experiment and its  signal
 discrimination power. 
One of us (V.A.) would like to thank M. Pallavicini for interesting
 and useful discussions about Borexino features and potentiality.
We thank also all those who sent us 
comments on early versions of this work.
We  acknowledge the  financial  support of 
 the Italian MURST, the  Spanish CYCIT  funding agencies 
 and the CERN Theoretical Division.
The numerical calculations have been performed in the computer farm of 
 the Milano University theoretical group (THEOS).

\newpage


\begin{table}[p]
\centering
\scalebox{1.}{
\begin{tabular}{|l|c|c|}
 \hline\vspace{0.1cm}
Experiment [Ref.] & $S_{SSM}$ & $S_{Data}/S_{SSM}\ (\pm 1\sigma)$
\\[0.1cm]\hline 
$SK$ (1258d)     \protect\cite{Fukuda:2001nk}  & 
$2.32\pm 0.03\pm 0.08 $ & $ 0.451\pm 0.011$   \\[0.1cm]
\CCSNO (240d, 8-20 MeV) \protect\cite{sno2001}  & 
$2.39\pm 0.34\pm 0.15$ &$0.347 \pm 0.029 $   \\[0.1cm]
$Cl$          \protect\cite{cl1999}  &  $2.56\pm 0.16\pm 0.16   $ & $0.332\pm 0.056$ \\[0.1cm]
$SAGE$        \protect\cite{sage1999,sage}  &  $67.2\pm 7.0 \pm 3.2  $ & $0.521\pm 0.067$ \\[0.1cm]
$GNO$-$GALLEX$  \protect\cite{gno2000,gallex}  &  $ 74.1\pm 6.7\pm 3.5 $ & $0.600\pm 0.067$  \\[0.1cm]
\hline
\end{tabular}
}
\caption{\small
Summary of data used in this work. 
The expected signal ($S_{SSM}$) and observed ratios $S_{Data}/S_{SSM}$
 with respect to the BPB2001 model are reported.
The $SK$ and \CCSNO rates are in \fluxunit units.
The $Cl$, $SAGE$ and $GNO$-$GALLEX$ measurements are in SNU units. 
In this work we use the combined results of $SAGE$ and $GNO$-$GALLEX$: 
 $S_{Ga}/S_{SSM}$ ($Ga\equiv SAGE$+$GALLEX$+$GNO$)=$0.579\pm 0.050 $.
The \boron total flux is taken from the BPB2001 model~\protect\cite{bpb2001}:
 $\phi_{\nu}(\boron)=5.05 (1^{+0.20}_{-0.16}) \times 10^6$ cm$^{-2}$ s$^{-1}$
.}
\label{tableratios}
\end{table}

\begin{table}[p]
 \begin{center}
  \scalebox{0.88}{
   \begin{tabular}{lr}
    \begin{tabular}{llcccc}
\hline
Solution & $ \Delta m^2$ & $\tan^2\theta$ &  
$\chi_{m}^2/n$ & g.o.f.\vphantom{K$^{K}_{K}$} \\
\hline
SMA          &  7.8 $\times 10^{-6} $ &  1.10 $\times 10^{-3} $&   1.1 &  29 \\ 
VAC          &  8.4 $\times 10^{-11} $&  0.27   &   1.3 &  26 \\ 
             &                        &  3.37   &   1.6 &  21 \\ 
LMA          &  1.2 $\times 10^{-5} $ &  0.34   &   1.9 &  17 \\
LOW          &  1.1 $\times 10^{-7} $ &  0.75   &   3.6 & 5.7 \\
\hline
      \end{tabular}
&
      \begin{tabular}{ccccc}
\hline
$SK$  & $Cl$ & $Ga$ & $SNO$ & $A^{DN}_{SK}$   \\
\hline
 0.5 &  0.8 &    0.2 &   2.5 &    1.3  \\ 
 1.9 &  1.4 &    0.8 &   0.2 &    1.3  \\
 1.7 &  1.4 &    0.9 &   0.3 &    1.3  \\
 0.0 &  1.3 &    0.2 &   0.7 &    1.1  \\
 2.0 &  1.9 &    1.0 &   2.5 &    0.3  \\
\hline
      \end{tabular}
      \end{tabular}
}
 \caption{\small
(Left) Best fit oscillation parameters:
 $ \Delta m^2 (\eV^2)$ and $\tan^2\theta$,
 from the analysis of global rates: $Cl$, $Ga$ and $SK$.
The $SNO$ measurement is not included (d.o.f.$=3-2$).
The significance level g.o.f. is in percentage.
The sampling error is $\sim 1\%$ for $\Delta m^2$ and 
  $\tan^2(\theta)$.
(Right) Minimization residuals in 1$\sigma$ units for diverse
 experimentally measured quantities: the global rates for the four
 experiments and  $SK$ global day-night 
 asymmetry ($A^{DN}$).
Only the first three quantities appear in the $\chi^2$ computation.
}
  \label{table1a}
 \end{center}
\end{table}

\begin{table}[p]
 \begin{center}
\scalebox{0.88}{
\begin{tabular}{lr}
      \begin{tabular}{ccccc}
\hline
Solution & $ \Delta m^2$ & $\tan^2\theta$ &  
$\chi_{m}^2/n$ & g.o.f.\vphantom{K$^{K}_{K}$} \\
\hline
LMA          &  3.1 $\times 10^{-5} $ &  0.40   &   1.6 &  21 \\
SMA          &  7.3 $\times 10^{-6} $ &  1.40 $\times 10^{-3} $&   2.4 &  8.8 \\ 
VAC          &  9.7 $\times 10^{-11} $&  0.26   &   3.1 &  4.6 \\ 
             &                        &  4.00   &   3.2 &  4.1\\ 
LOW          &  3.2 $\times 10^{-7} $ &  0.80   &   3.3 & 3.5 \\
\hline
      \end{tabular}
&
      \begin{tabular}{ccccc}
\hline
$SK$  & $Cl$ & $Ga$ & $SNO$ & $A^{DN}_{SK}$   \\
\hline
 1.4 &  0.9 &    1.5 &   1.3 &    1.6  \\ 
 1.0 &  1.5 &    0.4 &   0.9 &    1.3  \\
 3.8 &  2.3 &    1.1 &   1.5 &    1.3  \\
 4.1 &  2.5 &    1.3 &   1.9 &    1.3  \\
 2.7 &  1.2 &    3.1 &   1.7 &    0.9  \\
\hline
      \end{tabular}
      \end{tabular}
}
\caption{\small
Best fit oscillation parameters,
 $ \Delta m^2 (\eV^2)$ and $\tan^2\theta$, and minimization residuals
 (see explanation in Table~\ref{table1a}).
The analysis now includes the global rates for the four experiments:
 $Cl$, $Ga$, $SK$ and \CCSNO (d.o.f. $=4-2$).}
  \label{table1b}
 \end{center}
\end{table}

\begin{table}[p]
 \begin{center}
  \scalebox{0.88}{
   \begin{tabular}{lr}
    \begin{tabular}{ccccc}
\hline
Solution & $ \Delta m^2$ & $\tan^2\theta$ &  
$\chi_{m}^2/n$ & g.o.f.\vphantom{K$^{K}_{K}$} \\
\hline
LMA    &  5.4 $\times 10^{-5} $ & 0.36 &   0.8 &  82 \\
LOW    &  1.2 $\times 10^{-7} $ & 0.75 &   0.8 &  80 \\
VAC    &  2.7 $\times 10^{-10}$ & 2.70 &   0.8 &  80 \\ 
       &                        & 0.32 &   0.9 &  80\\ 
SMA    &  7.3 $\times 10^{-6} $ & 1.30 $\times 10^{-3} $&   1.1 &  40 \\ 
\hline
      \end{tabular}
&
      \begin{tabular}{ccccc}
\hline
$SK$  & $Cl$ & $Ga$ & $SNO$ & $A^{DN}_{SK}$   \\
\hline
 1.8 &  0.4 &    0.8 &   2.1 &    0.0  \\ 
 2.0 &  2.0 &    0.9 &   2.5 &    0.2  \\
 2.2 &  1.8 &    0.5 &   0.7 &    1.2  \\
 2.3 &  2.0 &    0.6 &   0.8 &    1.2  \\
  0.1 &  1.0 &    0.3 &   2.1 &    1.3  \\
\hline
     \end{tabular}
    \end{tabular}
}
   \caption{\small
Best fit oscillation parameters, $\Delta m^2 (\eV^2)$ and $\tan^2\theta$,
 and minimization residuals (see explanation in Table~\ref{table1a}).
The analysis now includes the global rates for three experiments
 $Cl$,$Ga$ and \CCSNO, and $SK$ day and night energy spectra.
There are four parameters: $\Delta m^2$, $\tan^2\theta$, $\alpha$ and 
 $\delta_{\rm cor}$.
We let the flux normalization $\alpha$ vary freely, so the d.o.f is
 $=41-4$ (see the text).
The sampling error due to the finite-grid is $\sim 1\%$ for
 $\Delta m^2_{min}$ and  $\tan^2(\theta)$.}
  \label{table2a}
 \end{center}
\end{table}

\begin{table}[p]
 \begin{center}
  \scalebox{0.88}{
   \begin{tabular}{lr}
      \begin{tabular}{llccc}
\hline
Solution & $ \Delta m^2$ & $\tan^2\theta$ &  
$\chi_{m}^2/n$ & g.o.f.\vphantom{K$^{K}_{K}$} \\
\hline
LMA    &  5.4 $\times 10^{-5} $ & 0.38 &   0.8 &  82 \\
LOW    &  7.5 $\times 10^{-8} $ & 0.84 &   0.9 &  72 \\
VAC    &  8.9 $\times 10^{-10}$ & 1.73 &   0.9 &  73 \\ 
       &                        & 0.48 &   0.9 &  63\\ 
SMA    &  7.3 $\times 10^{-6} $ & 1.30 $\times 10^{-3} $& 1.1 &  40 \\ 
\hline
      \end{tabular}
&
      \begin{tabular}{ccccc}
\hline
$SK$  & $Cl$ & $Ga$ & $SNO$ & $A^{DN}_{SK}$   \\
\hline
 1.6 &  0.5 &    0.9 &   1.8 &    0.0  \\ 
 2.7 &  2.3 &    0.8 &   2.9 &    0.7  \\
 2.4 &  1.9 &    0.5 &   0.7 &    1.3  \\
 2.4 &  2.2 &    1.4 &   1.6 &    1.3  \\
 0.1 &  1.0 &    0.3 &   2.1 &    1.3  \\
\hline
      \end{tabular}
      \end{tabular}
}
    \caption{\small
Same as Table~\ref{table2a} except that here the flux normalization
 $\alpha$ is constrained to vary with the SSM standard deviation and the
 d.o.f is $=41-3$ (see the text).}
  \label{table2b}
 \end{center}
\end{table}

\begin{table}[p]
 \begin{center}
  \scalebox{0.88}{
   \begin{tabular}{ccccc}
\hline
Solution & $ \Delta m^2 (\eV^2)$ & $\tan^2(\theta)$ &  
$S^{Bor}$  & $A^{DN}_{Bor}\  (\%)$ \\
\hline
LMA    &  5.4 $\times 10^{-5} $ & 0.38 &   0.63 &  -0.008   \\
LOW    &  7.5 $\times 10^{-8} $ & 0.84 &   0.61 &  -14.3 \\
VAC    &  8.9 $\times 10^{-10}$ & 1.73 &   0.64 &  +0.02  \\ 
       &                        & 0.48 &   0.65 &  -0.07  \\ 
SMA    &  7.3 $\times 10^{-6} $ & 1.30 $\times 10^{-3} $&  0.24 &  -0.02 \\ 
\hline
   \end{tabular}
}
      \caption{\small
Borexino experiment expectations for the normalized total signal 
 ($S^{Bor}$) and day-night asymmetry ($A^{DN}$) for all of the best-fit 
solutions appearing in Table~(\protect\ref{table2b}).}
      \label{table7}
   \end{center}
\end{table}

\end{document}